\pdfoutput=1

\documentclass[11pt,twoside,a4paper,cmspaper,final,collab]{cms-tdr}
\def\svnVersion{484861P}\def\cmsCernNoTag{CERN-EP-2018-233}\def\cmsCernDate{\today}\def\cmsMessage{Published in Physical Review Letters as \href{http://dx.doi.org/10.1103/Phys.Rev.Lett.121.241802}{\doi{10.1103/Phys.Rev.Lett.121.241802.}}}

\begin{document}\cmsNoteHeader{B2G-16-027}

\hyphenation{had-ron-i-za-tion}
\hyphenation{cal-or-i-me-ter}
\hyphenation{de-vices}
\RCS$HeadURL: svn+ssh://svn.cern.ch/reps/tdr2/papers/B2G-16-027/trunk/B2G-16-027.tex $
\RCS$Id: B2G-16-027.tex 482223 2018-11-23 16:27:11Z rkogler $

\ifthenelse{\boolean{cms@external}}{\providecommand{\cmsLeft}{upper\xspace}}{\providecommand{\cmsLeft}{left\xspace}}
\ifthenelse{\boolean{cms@external}}{\providecommand{\cmsRight}{lower\xspace}}{\providecommand{\cmsRight}{right\xspace}}
\providecommand{\NA}{\ensuremath{\text{---}}}
\ifthenelse{\boolean{cms@external}}{\providecommand{\CL}{C.L.\xspace}}{\providecommand{\CL}{CL\xspace}}

\newcommand{\ST}{\ensuremath{S_{\mathrm{T}}}\xspace}
\newcommand{\alphaST}{\ensuremath{\alpha(\ST)}\xspace}
\newcommand{\STlep}{\ensuremath{S_{\mathrm{T}}^{\text{lep}}}\xspace}
\newcommand{\Kstar}{\ensuremath{\PK^{(\ast)}}\xspace}
\newcommand{\RK}{\ensuremath{R_{\PK}}\xspace}
\newcommand{\RKstar}{\ensuremath{R_{\PK^{(\ast)}}}\xspace}
\newcommand{\Dstar}{\ensuremath{\PD^{(\ast)}}\xspace}
\newcommand{\RDstar}{\ensuremath{R_{\PD^{(\ast)}}}\xspace}
\newcommand{\amu}{\ensuremath{a_{\mu}}\xspace}
\newcommand{\LQ}{\ensuremath{\mathrm{LQ}}\xspace}
\newcommand{\LQs}{\ensuremath{\mathrm{LQs}}\xspace}
\newcommand{\MLQ}{\ensuremath{M_{\LQ}}\xspace}
\newcommand{\MLQrec}{\ensuremath{M_{\LQ}^{\text{rec}}}\xspace}
\newcommand{\LQtmu}{\ensuremath{\LQ\to\cPqt\mu}\xspace}
\newcommand{\LQttau}{\ensuremath{\LQ\to\cPqt\tau}\xspace}
\newcommand{\LQbnu}{\ensuremath{\LQ\to\cPqb\nu}\xspace}
\newcommand{\Btmu}{\ensuremath{\mathcal{B}(\LQtmu)}\xspace}
\newcommand{\Bttau}{\ensuremath{\mathcal{B}(\LQttau)}\xspace}
\newcommand{\Bbnu}{\ensuremath{\mathcal{B}(\LQbnu)}\xspace}

\cmsNoteHeader{B2G-16-027}
\title{Search for leptoquarks coupled to third-generation quarks in proton-proton collisions at \texorpdfstring{$\sqrt{s}=13\TeV$}{sqrt(s) = 13 TeV}}

\date{\today}

\abstract{
  Three of the most significant measured deviations from standard  model predictions, the enhanced decay rate for $\PB\to\Dstar\tau\nu$, hints of lepton universality violation in  $\PB\to\Kstar\ell\ell$ decays, and the anomalous magnetic moment of the muon, can be  explained by the existence of leptoquarks (\LQs) with large couplings to third-generation quarks and masses at the TeV scale. The existence of these states can be probed at the LHC in high energy proton-proton collisions. A novel search is presented for pair production of \LQs coupled to a top quark and a muon using data at a center-of-mass energy of 13\TeV, corresponding to an integrated luminosity of 35.9\fbinv, recorded by the CMS experiment. No deviation from the standard model prediction has been observed and scalar LQs decaying exclusively into $\cPqt\mu$ are excluded up to masses of 1420\GeV. The results of this search are combined with those from previous searches for \LQ decays into $\cPqt\tau$ and $\cPqb\nu$, which excluded scalar LQs below masses of 900 and 1080\GeV. Vector \LQs are excluded up to masses of 1190\GeV for all possible combinations of branching fractions to $\cPqt\mu$, $\cPqt\tau$ and $\cPqb\nu$. With this analysis, all relevant couplings of LQs with an electric charge of $-1/3$ to third-generation quarks are probed for the first time.
  }

\hypersetup{
pdfauthor={CMS Collaboration},%
pdftitle={Search for leptoquarks coupled to third-generation quarks in proton-proton collisions at sqrt(s) = 13 TeV},
pdfsubject={CMS},
pdfkeywords={CMS, physics, leptoquarks}}

\maketitle
The standard model of particle physics has been outstandingly successful in describing most fundamental physical phenomena. However, significant deviations from the predictions of the standard model (SM)
have been observed in measurements of rare decays of {\PB} mesons.
In particular, deviations have been seen in the values of the ratio \RDstar, defined as the ratio of the $\PB\to\Dstar\tau\nu$ branching fraction to the $\PB\to\Dstar\mu\nu$ branching fraction.
These deviations from the SM were first reported by the BaBar~\cite{Lees:2012xj,Lees:2013uzd} and Belle~\cite{Matyja:2007kt,Bozek:2010xy,Huschle:2015rga} Collaborations and have been confirmed by the LHCb Collaboration~\cite{Aaij:2015yra,Aaij:2017deq} with a combined significance of about four standard deviations~\cite{Amhis:2016xyh}.
The ratios of the branching fractions of $\PB\to\Kstar\mu\mu$ to $\PB\to\Kstar\Pe\Pe$, \RK and $R_{\PK^{\ast}}$, as measured by the LHCb Collaboration~\cite{Aaij:2014ora,Aaij:2014pli,Aaij:2015oid,Aaij:2017vbb}, show departures from lepton universality by 2.6 and 2.4 standard deviations, respectively.
The measurement of the muon anomalous magnetic moment \amu, one of the most precisely measured quantities in particle physics~\cite{Bennett:2006fi}, also deviates from the SM prediction by 3.5 standard deviations~\cite{Davier:2017zfy}. These anomalies are among the most significant deviations from the SM observed so far.

The existence of leptoquarks (\LQs) with masses at the TeV scale and
large couplings to third-generation quarks~\cite{Tanaka:2012nw,
Sakaki:2013bfa,Dorsner:2013tla,Dumont:2016xpj, Crivellin:2017zlb, Gripaios:2014tna, Bauer:2015knc, ColuccioLeskow:2016dox,
Becirevic:2017jtw,Hiller:2017bzc, Hiller:2018wbv}
has been proposed as a possible explanation for one, two, or all of these
deviations.
Leptoquarks are hypothetical particles that can decay to SM quarks and leptons.
They are triplets with respect to the strong interaction,
have fractional electric charge, and can be either scalar (spin 0) or vector (spin 1) particles.
Many extensions to the SM, among them grand unification~\cite{Pati:1974yy,PhysRevLett.32.438,Fritzsch:1974nn}, technicolor~\cite{Farhi:1980xs,PhysRevD.44.2678}, and compositeness models~\cite{Schrempp:1984nj,Gripaios:2009dq}, predict the existence of these particles.
The effective Buchm\"uller--R\"uckl--Wyler model~\cite{BRW} incorporates the assumption that \LQ interactions with SM fermions are renormalizable and gauge invariant, leading to restrictions on the allowed quantum numbers of \LQs~\cite{Acosta:1999ws}. 
Depending on its quantum numbers and the coupling structure, a given \LQ can decay to any one of a number of different combinations of SM fermions.
The couplings of \LQs to leptons and quarks of different generations introduce flavor changing neutral currents that may be observable in precision measurements~\cite{Dorsner:2016wpm}.
While simultaneous couplings to the first and second generations are tightly constrained by experimental data, the bounds are weaker for couplings to the second and third generation, thus allowing the existence of leptoquarks with non-diagonal couplings in the generation matrix~\cite{Crivellin:2017zlb, Hiller:2017bzc, Diaz:2017lit}.

Collider searches for \LQs with decays to third-generation quarks have
been performed in the decay channels \LQttau,
$\LQ\to\cPqb\tau$, and \LQbnu at
$\sqrt{s}=8\TeV$~\cite{Khachatryan:2015bsa,Abazov:2007bsa,Aaltonen:2007rb,ATLAS:2013oea,
  Chatrchyan:2012st,Khachatryan:2014ura,Khachatryan:2015wza,Aad:2015caa} and recently at
  $\sqrt{s}=13\TeV$~\cite{Khachatryan:2016jqo,
  Sirunyan:2017yrk, Sirunyan:2017kqq, Sirunyan:2018nkj, Sirunyan:2018kzh}.
We present the first search for the
pair production of \LQs with decays
to a top quark and a muon, \LQtmu, a decay mode
that is essential to explain the anomalies in \amu and
\RKstar~\cite{Gripaios:2014tna,Bauer:2015knc,ColuccioLeskow:2016dox,
Becirevic:2017jtw,Hiller:2017bzc, Crivellin:2017zlb, Hiller:2018wbv}.
This search is combined with previous searches that target other decay modes~\cite{Sirunyan:2018nkj, Sirunyan:2018kzh}.
The combination provides sensitivity to all relevant couplings of \LQs with an electric charge of $-1/3$ to third-generation quarks.

At the CERN LHC, pair production of LQs is possible via gluon-gluon  fusion or quark-antiquark annihilation, allowing direct searches to be performed.
Single LQ production via quark-gluon scattering is subdominant for LQs coupled to heavy quarks, as it requires a heavy quark in the initial state. The pair production cross section depends on the mass of the scalar LQ and is known at next-to-leading order (NLO) precision~\cite{Kramer:2004df}. The pair production cross section for vector LQs has been calculated at leading order (LO)~\cite{Dorsner:2018ynv} and is much larger than the scalar LQ cross section. The cross section for vector LQs depends on an additional parameter $\kappa$, which is a dimensionless coupling and takes a value of $\kappa=1$ in the Yang-Mills case and $\kappa=0$ in the minimal coupling case.

The central feature of the CMS apparatus is a superconducting solenoid
of 6\unit{m} internal diameter, providing a magnetic field of
3.8\unit{T}. Within the solenoid volume are a silicon pixel and strip
tracker, a lead tungstate crystal electromagnetic calorimeter,
and a brass and scintillator hadron calorimeter, each composed
of a barrel and two endcap sections. Forward calorimeters extend the
pseudorapidity ($\eta$) coverage provided by the barrel and endcap
detectors. Muons are detected in gas-ionization chambers embedded in
the steel flux-return yoke outside the solenoid. A more detailed
description of the CMS detector, together with a definition of the
coordinate system used and the relevant kinematic variables, can be
found in Ref.~\cite{CMS}.

This analysis uses data recorded by the CMS detector in {\Pp\Pp} collisions at a center-of-mass energy of 13\TeV in 2016.
Online, potential signal events are required to pass a single-muon trigger that selects isolated muon candidates with transverse momentum $\pt>24$\GeV~\cite{Khachatryan:2016bia}.
Data recorded by single electron triggers are used in background-enriched control regions (CRs).
The data correspond to an integrated luminosity of 35.9\fbinv.

Signal events of pair-produced \LQs with prompt decays to $\cPqt\mu$ are simulated
with the \PYTHIA 8.205~\cite{Sjostrand:2006za, Sjostrand:2014zea} Monte Carlo program at LO for mass values ranging from 200 to 2000\GeV.
The \POWHEG~\cite{Nason:2004rx,Frixione:2007vw,Alioli:2010xd,Frixione:2007nw,
Melia:2011tj,Nason:2013ydw,Alioli:2009je,Re:2010bp}
v1 generator is used to simulate background events resulting from the production of single top quarks in the \cPqt\PW-channel at NLO. The \POWHEG v2 generator is used for single top production in the $t$-channel and for simulating \ttbar production at NLO. Single top quark production in the $s$-channel, \ttbar production in association with a heavy gauge boson (\ttbar+ V), and the production of a {\PW} boson with additional jet radiation are simulated with \MGvATNLO (v2.2.2)~\cite{Alwall:2014hca} at NLO.
Events from Drell--Yan (DY) production with additional jet radiation are simulated
with \MGvATNLO at LO and an NLO $K$ factor is applied to the LO DY+jets production cross section. The simulation of the production
of two heavy gauge bosons with additional jet radiation is performed at NLO
with \MGvATNLO and \POWHEG v2.
Events in which jets are produced through the strong interaction only, referred to as quantum chromodynamic multijet events, are simulated with \PYTHIA at LO.

Parton showers in the simulated {\PW} boson production
events and DY events with additional jet radiation
are matched to the matrix element calculation with the
FxFx~\cite{Frederix:2012ps} and
MLM~\cite{Alwall:2007fs} algorithms, respectively. The parton shower
and hadronization process is simulated with \PYTHIA.
The NNPDF3.0~\cite{Ball:2014uwa} parton distribution functions
(PDFs) at LO and NLO are used for processes simulated at LO and NLO, respectively. The underlying event tune
CUETP8M2T4~\cite{CMS-PAS-TOP-16-021} is used for the simulation of \ttbar and single top
quark production via the $t$-channel, all other
processes are generated using
CUETP8M1~\cite{Khachatryan:2015pea,Skands:2014pea}.
All simulated event samples include the simulation of additional
inelastic {\Pp\Pp} interactions within the same or adjacent
bunch crossings (pileup).  The detector response is simulated with the
\GEANTfour package~\cite{Agostinelli:2002hh, Allison:2006ve}. Simulated
events are processed through the software chain used for
collision data and are reweighted to match the observed distribution
of the number of pileup interactions in data.

The CMS experiment uses a particle-flow (PF) event reconstruction
algorithm~\cite{Sirunyan:2017ulk}, which makes use of an optimized
combination of information from the various elements of the CMS detector.
The reconstructed vertex with the largest value of summed physics
object $\pt^2$ is taken to be the primary {\Pp\Pp} interaction vertex.
The physics objects here are the objects returned by a jet finding
algorithm~\cite{Cacciari:2008gp,Cacciari:2011ma} applied to all charged
tracks associated with the vertex, plus the associated missing transverse momentum, taken as the negative vector \pt sum of those jets. More details are given in Ref.~\cite{CMS-TDR-15-02}.
All detected particles are reconstructed either as electrons, muons, photons, charged
hadrons, or neutral hadrons.
In this analysis, electrons and muons are required to have $\pt\geq 30\GeV$, $\abs{\eta}\leq 2.4$, and to be isolated.
The isolation~\cite{Sirunyan:2018fpa, Khachatryan:2015iwa} is defined as the summed \pt of all neutral particles and charged hadrons in a cone with radius $\Delta R = \sqrt{\smash[b]{(\Delta\eta)^2+(\Delta\phi)^2}}$, $\phi$ being the azimuthal angle in radians, of 0.4 (for muons) or 0.3 (for electrons) around the lepton. The sum is corrected for the contribution of neutral pileup inside the cone.
Jets are clustered from charged and neutral PF candidates using the anti-\kt jet-clustering algorithm~\cite{Cacciari:2008gp, Cacciari:2011ma} with a distance parameter of 0.4. Charged PF candidates originating from vertices other than the primary vertex are not clustered.
A jet energy correction (JEC) is applied~\cite{Khachatryan:2016kdb} to account for remaining contributions arising from a different vertex than the primary one as well as for non-uniformity of the jet response in $\eta$ and non-linearity in $\pt$. Finally, a correction is applied to account for the residual differences in the jet response between data and simulated events.
The jet energy resolution (JER) in simulated events is smeared to match the wider resolution in data.
All jets are required to have $\pt\geq 30$\GeV and $\abs{\eta}\leq 2.4$.
The combined secondary vertex v2~\cite{CSVv2} algorithm is used to
identify jets originating from bottom quarks (\cPqb-tagged jets). The loose working point is chosen, which has an efficiency of about 90\% and a mistag rate of approximately 10\%.
The missing transverse momentum, \ptmiss, is calculated as the
magnitude of the negative vectorial \pt sum of all PF
candidates in an event. Both the jet energy scale and resolution corrections
are propagated to the calculation of \ptmiss.

Offline, events are required to contain at least two muons and at least
two jets, of which at least one must be \cPqb-tagged. By requiring the invariant mass of each
pair of muons in an event to exceed the \PZ boson mass by at least 20\GeV, events arising from the production of a \PZ boson with additional jet radiation are suppressed.
As the decay of heavy LQs is expected to produce highly energetic leptons and jets, minimum values of \STlep and \ST of 200 and 350\GeV are required, respectively. Here, \STlep is the scalar \pt sum of all selected muons and electrons and \ST is defined as the scalar sum of \STlep, \ptmiss, and the \pt of all selected jets.
The phase space region resulting from these selection criteria is referred to as the signal region (SR) in the following.

Two categories of events are defined, based on the number of muons or electrons. If at least three such charged leptons are present, of which at least two are muons, and at least one pair of muons with opposite electric charge is found, the event falls into category A. Category B contains all remaining events in the SR.
In category A, the LQ mass \MLQrec is reconstructed under the assumption that one of the top quarks decays into the leptonic final state with a muon or an electron (leptonic top) and the other one decays into the hadronic final state (hadronic top). The distribution of \MLQrec is used for the final statistical analysis in this category, while the distribution of \ST is used for this purpose in category B.

For each event, the leptonic top quark candidates are constructed from permutations of one or more of the seven \pt-leading jets, one of the three \pt-leading muons or the \pt-leading electron, and \ptmiss. The hadronic top quark candidates are constructed using all permutations of jets not assigned to the leptonic top quark. The LQ candidates are assembled from top quark candidates and the two \pt-leading muons that have not been associated to the leptonic top quark. 
The muon charge is used when assigning it to one of the top quark candidates.
In events with more than two muons, all possible permutations of muons are considered.
A $\chi^2$ variable that takes into account the invariant mass of each top quark candidate and the relative mass difference between the two LQ candidates is then used to select the best pair of LQ candidates for each event.
Events with four leptons, which could originate from dileptonic \ttbar decays, are included in category A and contribute to the signal efficiency. 
In order to provide a more accurate SM background prediction in category A, which contains a minimum of three charged leptons, the misidentification rate of electrons and muons is measured using jets in a DY+jets enriched CR in data. 
The CR is defined by selecting two muons with an invariant mass close to the Z boson mass, and the misidentification rate is measured on events where a jet is misidentified as a third lepton. 
The resulting data-to-simulation correction factors are applied to simulation for each misidentified charged lepton in a given event in the SR. 
The effect of charge misidentification on the analysis was found to be negligibly small.

The contributions from the dominant SM backgrounds in category B,
the production of \ttbar and DY+jets events, are estimated simultaneously
in a data-driven procedure. A CR similar to the SR is defined
by requiring a minimum of two electrons without additional muons.
The invariant mass of any pair of electrons must be at least 20\GeV above the \PZ boson mass and all other SR requirements have to be fulfilled for the selected events.
We correct for small differences in the distribution of \ST between the
SR and the CR with an extrapolation function \alphaST,
which is derived from simulated \ttbar and DY+jets events by fitting both \ST distributions with an empirical functional form to obtain smoothed distributions that are then used to compute the ratio.
The number of data events in the CR, after all simulated minor backgrounds have been subtracted, is multiplied by \alphaST to extrapolate into the SR.
Using the ratio of the fitted functions results in a significantly smaller impact of systematic uncertainties on the estimated backgrounds.

Various uncertainties affecting the rate and the shape of the signal and background contributions are taken into account. In general, uncertainties in this analysis are treated similarly to those in Ref.~\cite{Sirunyan:2018nkj}.
For the background in category A, the uncertainties in the renormalization and factorization scales as well as the uncertainties in the lepton misidentification rates are dominant.
In category B the major backgrounds are derived from data. Lepton efficiencies and the background extrapolation procedure are the most important sources of uncertainties for these backgrounds. Uncertainties in the renormalization and factorization scales and in those associated with the choice of PDFs~\cite{Ball:2014uwa,Butterworth:2015oua} used to simulate the events dominate for the minor backgrounds.
The signal in both categories is most affected by the uncertainties in lepton and \cPqb-tagging~\cite{CSVv2} efficiencies. Other uncertainties considered are related to SM cross sections~\cite{TT,W/Z,ST1, ST2, ST3,DB1,DB2, DB3,TTV}, the integrated luminosity~\cite{LUMI}, JEC/JER~\cite{Khachatryan:2016kdb}, and the pileup reweighting~\cite{Sirunyan:2018nqx}.

\begin{figure}
\centering
\includegraphics[width=0.49\textwidth]{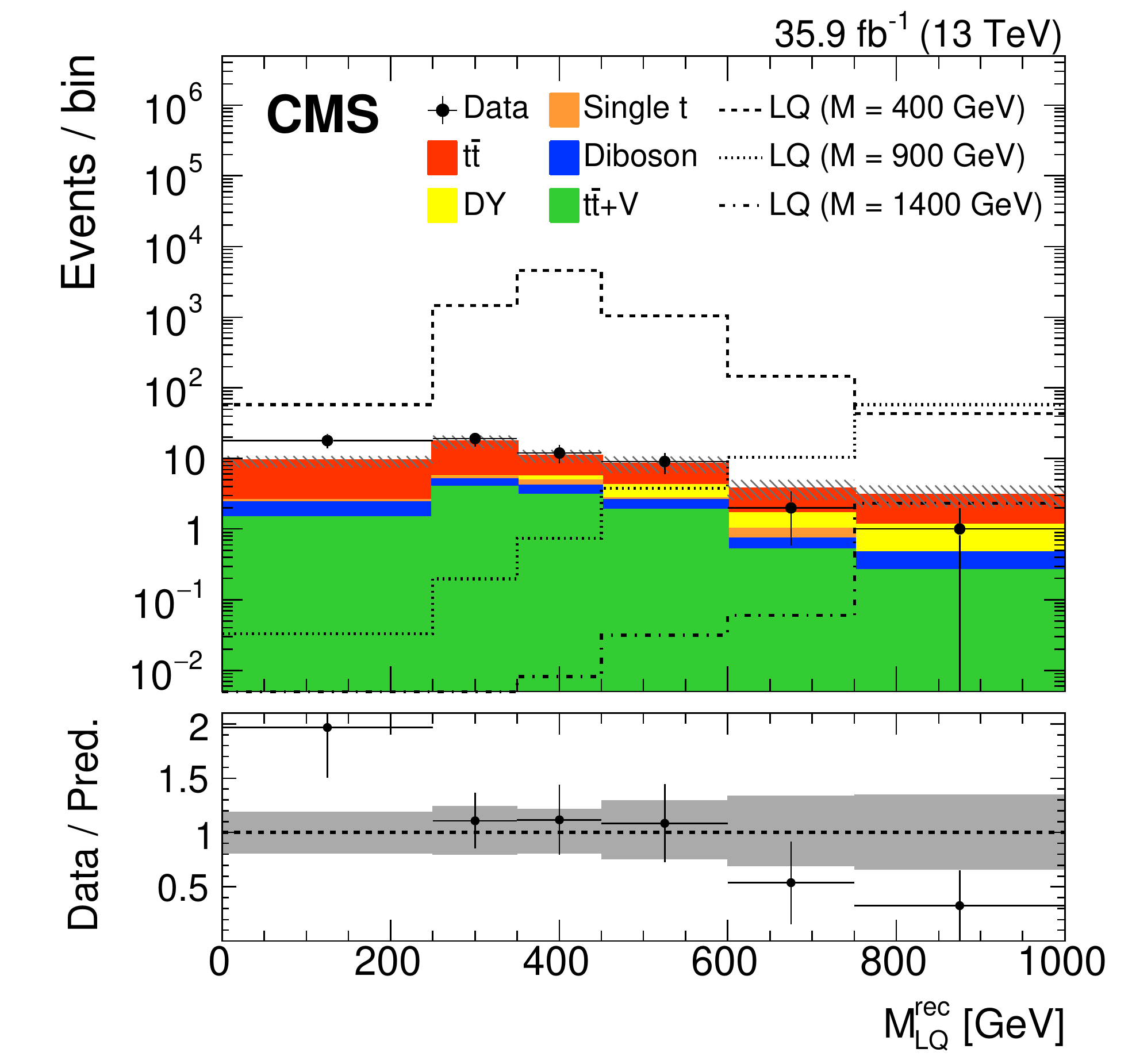}
\includegraphics[width=0.49\textwidth]{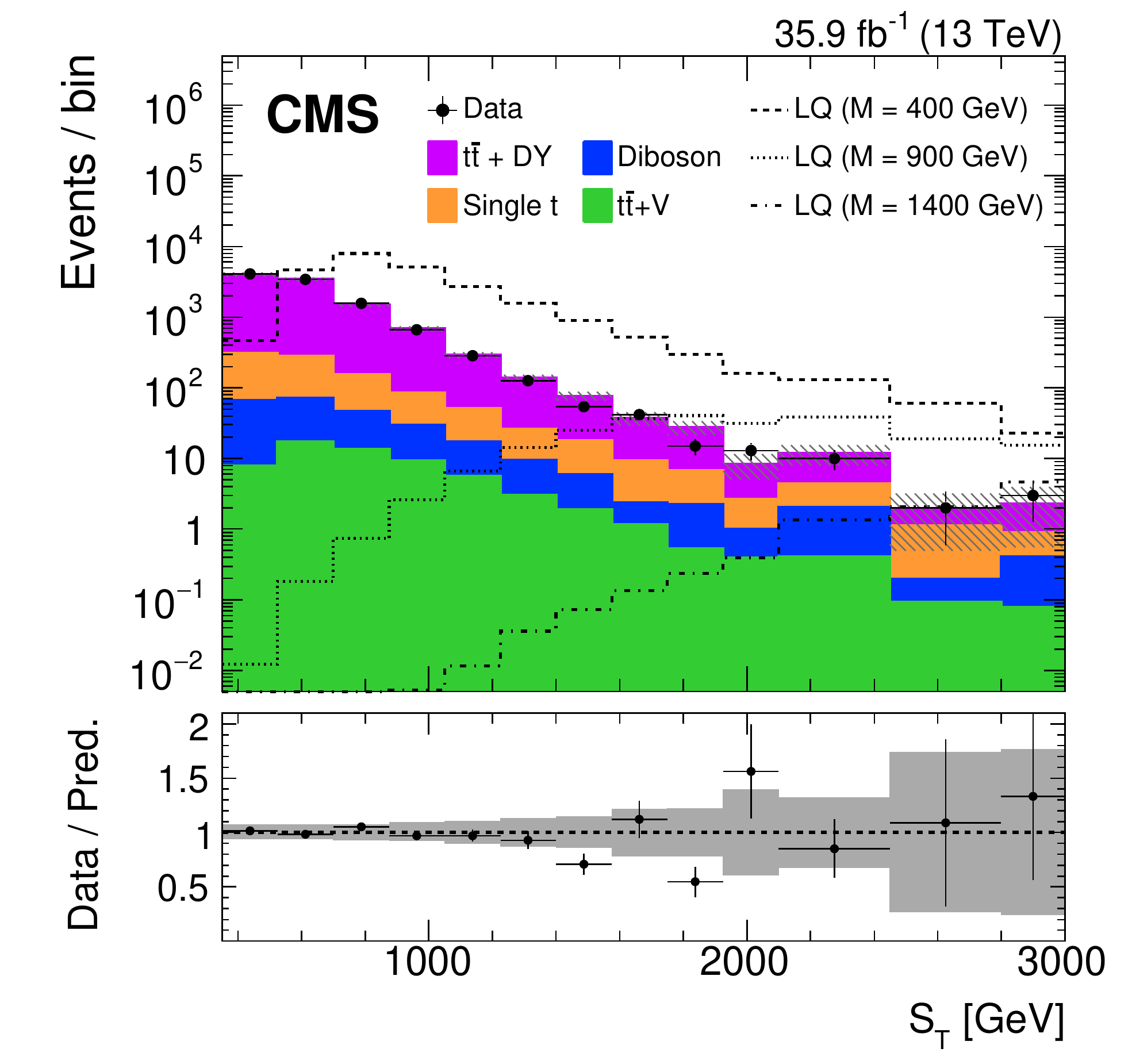}
\caption{Distributions for \MLQrec (category A, \cmsLeft) and \ST (category B, \cmsRight) after applying the full selection and estimating the \ttbar and DY+jets background contributions from data in category B. All backgrounds are normalized according to the post-fit nuisance parameters based on the corresponding SM cross sections. In the upper panels, the hatched areas correspond to the total uncertainty. In the lower panels, the gray bands indicate the total uncertainty.}
\label{fig:FinalDist}
\end{figure}

The \textsc{theta} software package~\cite{theta} is used to perform a maximum-likelihood template fit to the binned \MLQrec and \ST distributions for the background and to extract the cross section of a potential signal.
The statistical uncertainties in the SM backgrounds and the signal, as well as all systematic uncertainties, are taken into account as nuisance parameters in the fit. The uncertainty in the luminosity is assigned a log-normal prior distribution, for all other systematic uncertainties a Gaussian prior is used. The statistical uncertainty in the predicted background and the signal is taken into account by defining one additional nuisance parameter with a Gaussian distribution for each bin. A flat prior distribution is assumed for the signal cross section.
The data are found to be compatible with the SM prediction in both categories. The distributions of \MLQrec and \ST after the background-only fit are shown in Fig.~\ref{fig:FinalDist}.
A Bayesian method~\cite{theta,soton46376,CowanPDGStat} is used to set upper limits at 95\% confidence level (\CL) on the cross section for pair production of LQs decaying into a top quark and a muon.
Pseudo-experiments are performed to determine the median along with the regions expected to contain 68 and 95\% of the distribution of limits under the background-only hypothesis.

\begin{figure}
\centering
\includegraphics[width = 0.49\textwidth]{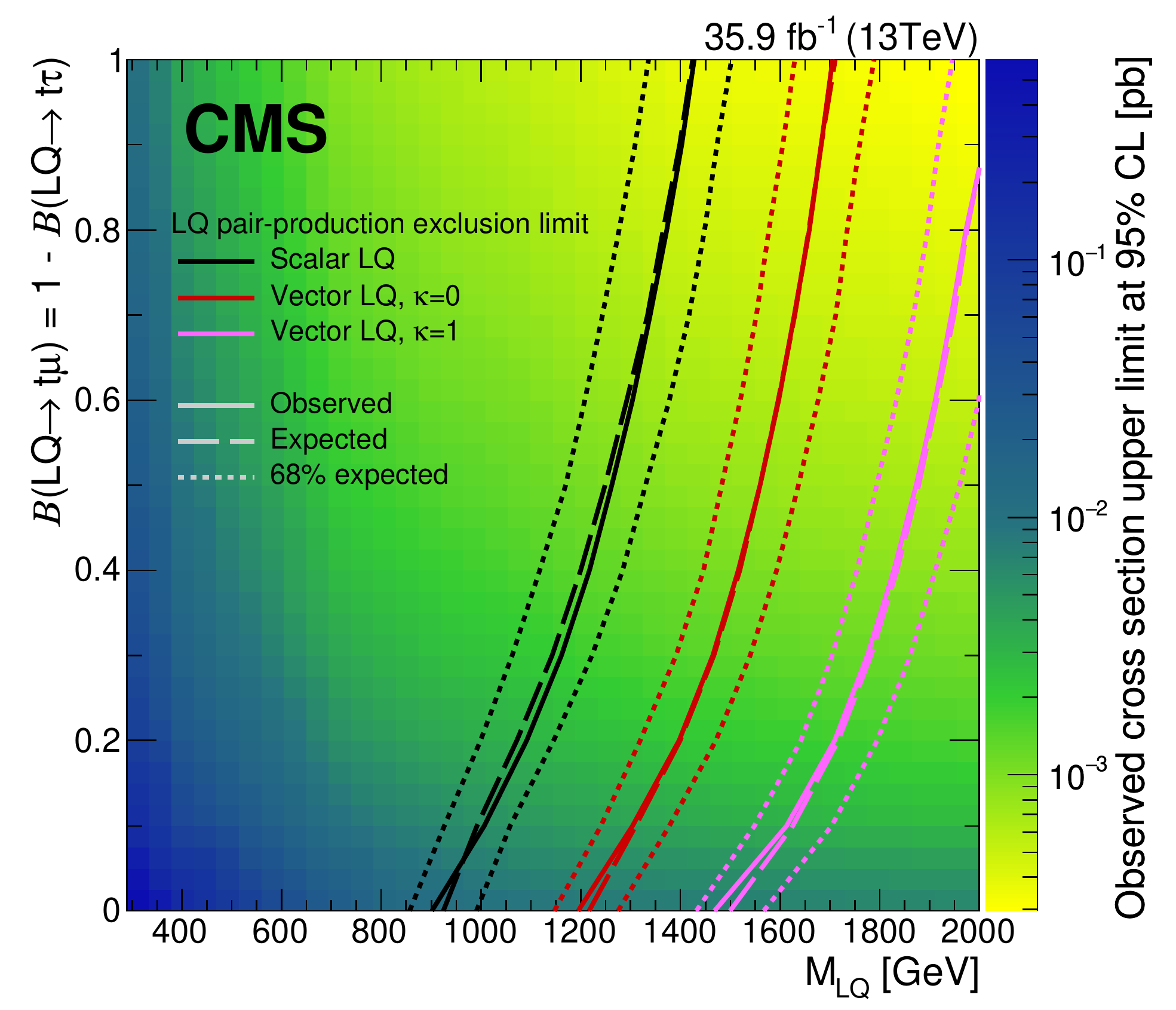}
\includegraphics[width = 0.49\textwidth]{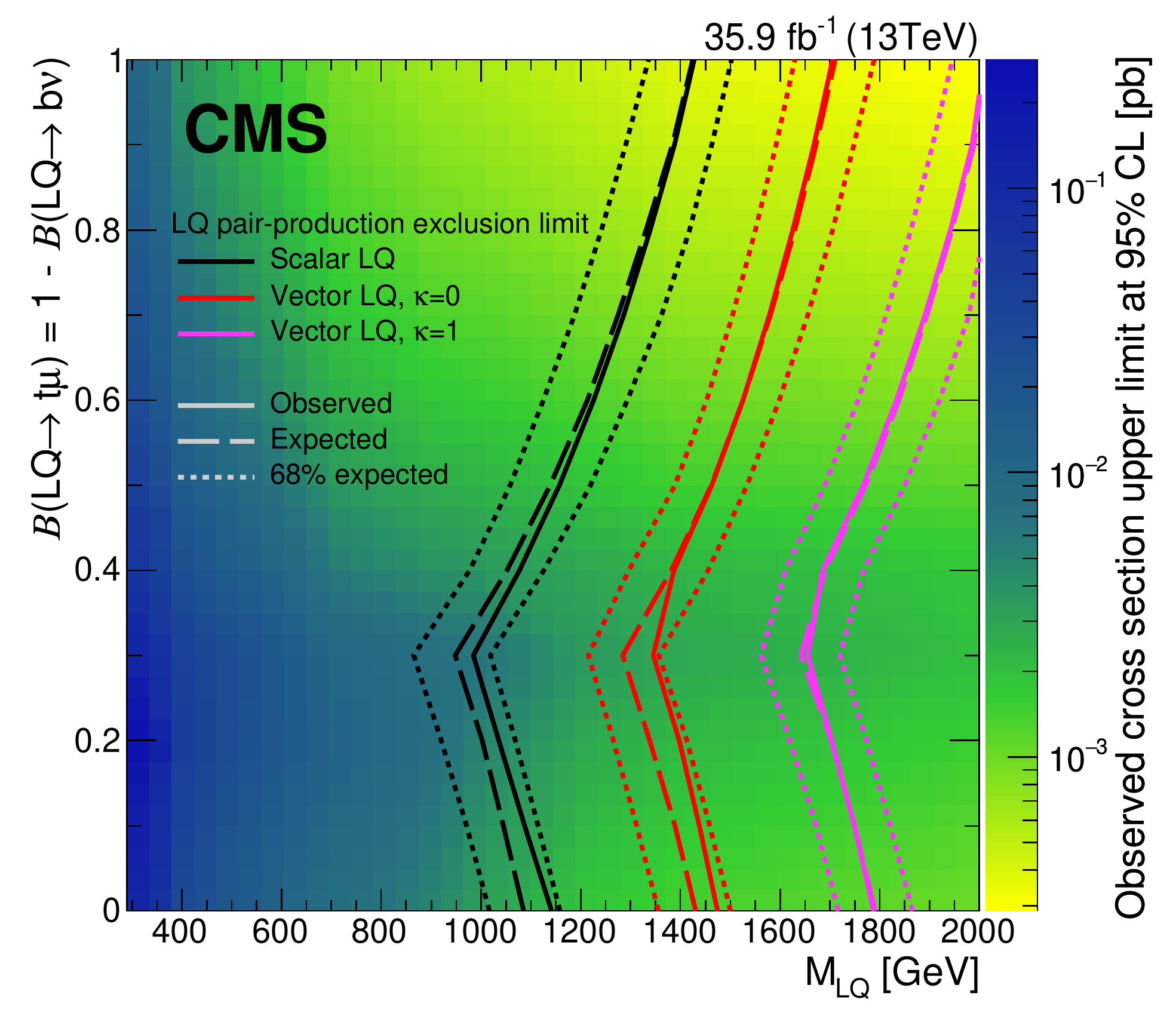}
\caption{Observed upper limits on the production cross section for pair production of
LQs decaying into a top quark and a muon or a $\tau$ lepton (\cmsLeft) and LQs decaying into a top quark and a muon or into a bottom quark and a neutrino (\cmsRight) at 95\% \CL in the \MLQ--\Btmu plane.
The lines show the lower mass exclusion limits for scalar (black) and vector (colored) LQs. They are derived by using the prediction for the scalar and vector LQ signal calculated at NLO~\cite{Kramer:2004df} and LO~\cite{Dorsner:2018ynv}, respectively.}
\label{fig:Limits2d}
\end{figure}

Pair-produced scalar LQs decaying exclusively into a top quark and a muon, $\Btmu = 1$, are excluded 
at 95\% \CL for \LQ masses up to 1420\GeV, exceeding the best previous limit, 
obtained from a reinterpretation~\cite{Diaz:2017lit} 
of a search for supersymmetry~\cite{Sirunyan:2017hvp}, by more than 600\GeV.
These results are combined with results from the \LQttau~\cite{Sirunyan:2018nkj} and \LQbnu~\cite{Sirunyan:2018kzh} decay channels to set exclusions limits in the plane of \MLQ and \Btmu.
Figure~\ref{fig:Limits2d} presents upper limits on the product of the production cross section and the branching fraction squared for $\Btmu = 1-\Bttau$ (\cmsLeft) and $\Btmu = 1-\Bbnu$ (\cmsRight). The values for $\Btmu = 0$ correspond to the results of the search for pair-produced LQs in the \LQttau decay channel (\cmsLeft) and the search for pair-produced LQs in the \LQbnu channel (\cmsRight). These analyses excluded pair-produced scalar \LQs in the targeted decay channels up to $\MLQ = 900 $ and 1080\GeV, respectively.
In the \cmsLeft (\cmsRight) part of Fig.~\ref{fig:Limits2d} the sensitivity is driven by the present analysis for values of $\Btmu > 0.1 (0.3)$ and by the $\LQttau (\cPqb \nu)$ search for smaller values.
Scalar \LQs decaying into a top quark and either a muon or a $\tau$ lepton are excluded below masses of 900\GeV for all values of \Btmu, whereas LQs
decaying either into a top quark and a muon or into a bottom quark and a neutrino are excluded up to $\MLQ = 980\GeV$.
The simulated samples of scalar LQ pair production are also used to derive mass exclusion limits for pair-produced vector LQs, as the acceptance for both types of LQs is similar. The lower limit of excluded vector LQ masses is shown in Fig.~\ref{fig:Limits2d} for the two coupling cases $\kappa=1$ and $\kappa=0$. Vector LQs are excluded up to masses of 1190\GeV for all values of \Btmu and $\kappa$ considered.

In summary, this analysis represents the first search for leptoquarks (\LQs) decaying to top quarks and muons, reaching \LQ masses of $\mathcal{O}(1\TeV)$ and placing direct constraints on the corresponding \LQ coupling, thus probing the region of interest of models including \LQs. With this result, all relevant couplings of \LQs with an electric charge of $-1/3$ to third-generation quarks are examined for the first time.

\begin{acknowledgments}
\hyphenation{Bundes-ministerium Forschungs-gemeinschaft Forschungs-zentren} We congratulate our colleagues in the CERN accelerator departments for the excellent performance of the LHC and thank the technical and administrative staffs at CERN and at other CMS institutes for their contributions to the success of the CMS effort. In addition, we gratefully acknowledge the computing centers and personnel of the Worldwide LHC Computing Grid for delivering so effectively the computing infrastructure essential to our analyses. Finally, we acknowledge the enduring support for the construction and operation of the LHC and the CMS detector provided by the following funding agencies: BMWFW and FWF (Austria); FNRS and FWO (Belgium); CNPq, CAPES, FAPERJ, and FAPESP (Brazil); MES (Bulgaria); CERN; CAS, MoST, and NSFC (China); COLCIENCIAS (Colombia); MSES and CSF (Croatia); RPF (Cyprus); SENESCYT (Ecuador); MoER, ERC IUT, and ERDF (Estonia); Academy of Finland, MEC, and HIP (Finland); CEA and CNRS/IN2P3 (France); BMBF, DFG, and HGF (Germany); GSRT (Greece); OTKA and NIH (Hungary); DAE and DST (India); IPM (Iran); SFI (Ireland); INFN (Italy); MSIP and NRF (Republic of Korea); LAS (Lithuania); MOE and UM (Malaysia); BUAP, CINVESTAV, CONACYT, LNS, SEP, and UASLP-FAI (Mexico); MBIE (New Zealand); PAEC (Pakistan); MSHE and NSC (Poland); FCT (Portugal); JINR (Dubna); MON, RosAtom, RAS, RFBR and RAEP (Russia); MESTD (Serbia); SEIDI, CPAN, PCTI and FEDER (Spain); Swiss Funding Agencies (Switzerland); MST (Taipei); ThEPCenter, IPST, STAR, and NSTDA (Thailand); TUBITAK and TAEK (Turkey); NASU and SFFR (Ukraine); STFC (United Kingdom); DOE and NSF (USA).
\end{acknowledgments}

\bibliography{auto_generated}

\cleardoublepage \appendix\section{The CMS Collaboration \label{app:collab}}\begin{sloppypar}\hyphenpenalty=5000\widowpenalty=500\clubpenalty=5000\vskip\cmsinstskip
\textbf{Yerevan Physics Institute, Yerevan, Armenia}\\*[0pt]
A.M.~Sirunyan, A.~Tumasyan
\vskip\cmsinstskip
\textbf{Institut f\"{u}r Hochenergiephysik, Wien, Austria}\\*[0pt]
W.~Adam, F.~Ambrogi, E.~Asilar, T.~Bergauer, J.~Brandstetter, M.~Dragicevic, J.~Er\"{o}, A.~Escalante~Del~Valle, M.~Flechl, R.~Fr\"{u}hwirth\cmsAuthorMark{1}, V.M.~Ghete, J.~Hrubec, M.~Jeitler\cmsAuthorMark{1}, N.~Krammer, I.~Kr\"{a}tschmer, D.~Liko, T.~Madlener, I.~Mikulec, N.~Rad, H.~Rohringer, J.~Schieck\cmsAuthorMark{1}, R.~Sch\"{o}fbeck, M.~Spanring, D.~Spitzbart, A.~Taurok, W.~Waltenberger, J.~Wittmann, C.-E.~Wulz\cmsAuthorMark{1}, M.~Zarucki
% ...
\vskip\cmsinstskip
% ...
1:  Also at Vienna University of Technology, Vienna, Austria\\
% ...
\end{sloppypar}
\end{document}